\documentclass[conference]{IEEEtran}
\IEEEoverridecommandlockouts
\usepackage{cite}
\usepackage{amsmath,amssymb,amsfonts}
\usepackage{algorithmic}
\usepackage{graphicx}
\usepackage{textcomp}
\usepackage{xcolor}
\def\BibTeX{{\rm B\kern-.05em{\sc i\kern-.025em b}\kern-.08em
    T\kern-.1667em\lower.7ex\hbox{E}\kern-.125emX}}

\usepackage[OT1]{fontenc}

\usepackage[english]{babel}

\usepackage[utf8]{inputenc}

\usepackage[sc]{mathpazo}

\usepackage{mathrsfs}

\usepackage[amsmath,thmmarks]{ntheorem}

\usepackage{soul}
\usepackage{acronym}
\usepackage{nomencl}
\usepackage{pdfpages}

\renewcommand{\epsilon}{\ensuremath\varepsilon}

\renewcommand{\phi}{\ensuremath{\varphi}}

\newcommand{\cM}{\mathcal{M}}

\newcommand{\cS}{\mathcal{S}}	%

\def\thmspace{0.2em}

\newtheorem{definition}{\hspace{\thmspace}{\bf Definition}\!}

\usepackage[linkcolor=black,colorlinks=true,citecolor=black,filecolor=black]{hyperref}

\usepackage{mathtools}

\usepackage{array}

\usepackage{bm} 

\usepackage{nomencl}
\usepackage{etoolbox}
\usepackage{wrapfig}

\usepackage{eurosym}

\begin{document}

\title{Differentially-Private Heat and Electricity Markets Coordination\\
\thanks{This work is partially funded by the NCCR Automation.}
}

\author{\IEEEauthorblockN{Lesia Mitridati, Emma Romei, Gabriela Hug}
\IEEEauthorblockA{\textit{Institute for Power Systems \& High Voltage Technology} \\
\textit{ETH Zurich}\\
Zürich, Switzerland \\
\{mitridati,eromei,hug\}@eeh.ee.ethz.ch}
\and
\IEEEauthorblockN{Ferdinando Fioretto}
\IEEEauthorblockA{\textit{Dept. of Electrical Engineering and Computer Science} \\
\textit{Syracuse University}\\
Syracuse, USA \\
ffiorett@syr.edu}
}

\maketitle

\begin{abstract}
Sector coordination between heat and electricity systems has been identified has an energy-efficient and cost-effective way to transition towards a more sustainable energy system.
However, the coordination of sequential markets relies on the exchange of sensitive information between the market operators, namely time series of consumers' loads. To address the privacy concerns arising from this exchange, this paper introduces a novel privacy-preserving Stackelberg mechanism (\emph{w}-PPSM) which generates differentially-private data streams with high fidelity. The proposed \emph{w}-PPSM enforces the \textit{feasibility} and \textit{fidelity} of the privacy-preserving data with respect to the original problem through a post-processing phase in order to achieve a close-to-optimal coordination between the markets. Multiple numerical simulations in a realistic energy system demonstrate the effectiveness of the \emph{w}-PPSM, which achieves up to two orders of magnitude reduction in the cost of privacy compared to a traditional differentially-private mechanism.
\end{abstract}


\section{Introduction}

The development of market-based coordination mechanisms for heat and electricity systems has been identified as a crucial step towards an energy-efficient, cost-effective, and sustainable energy system \cite{heatroadmap,pinson2017towards}. Recent advances in the literature have modelled the coordination between \textit{sequential} and \textit{interdependent} markets as a
\textit{Stackelberg game} \cite{mitridati2020heat,mitridati2019bid,byeon2019unit}. In particular, the electricity-aware heat market (EAHM) developed in \cite{mitridati2020heat} provides a market-based mechanism for the coordination of heat and electricity systems. This market framework is modelled as a bilevel optimization problem and relies on the sharing of information between the electricity and heat market operators to achieve an optimal coordination.

Despite recent regulatory changes encouraging information exchange between system operators \cite{FERCreport}, users in the electricity market may be reluctant to exchange some information with the heat market operator due to privacy concerns. Revealing this sensitive data may provide a competitive advantage over other strategic
agents, reveal identifying personal information, induce financial losses and security risks for the users, and even
benefit external attackers \cite{lisovich2008privacy,maharjan2013dependable}. In particular, in the EAHM developed in \cite{mitridati2020heat} and used in this paper as a target application, we consider that the hourly electricity loads of individual consumers represent a sensitive data stream to be obfuscated before releasing to the heat market operator.

To address this privacy issue, \emph{Differential Privacy} (DP) has emerged as a robust privacy framework for multiple applications \cite{dwork:13}. DP relies on the injection
of carefully calibrated noise to protect the disclosure of the individuals' data, while allowing to extract information about the population. This framework can thus be used to \emph{obfuscate} the sensitive data exchanged between the electricity and heat market operators in the EAHM. In particular, the $w$-privacy framework introduced in \cite{W_levelP} provides an interesting framework to obfuscate time series of hourly data, such as electricity loads, within a predefined time window.
However, the obfuscation of highly correlated and high-dimensional streams of data is particularly challenging due to the high level of noise required to maintain privacy goals \cite{OptStream}. When obfuscated data is used as input to optimization problems with strong techno-economic constraints, such as market clearing problems in energy systems, it may lead to severe fidelity and feasibility issues.
To address this issue the authors in \cite{mak2019privacy} developed an optimization-based fidelity-recovery phase to classic DP mechanisms. This approach has been adapted to the exchange of information in Stackelberg games, and applied to the coordination of electricity and natural gas markets in \cite{fioretto2021differential}. However, these recent advances in the literature are limited to classic definitions of DP. To the best of our knowledge, there is no existing mechanism to share differentially-private data streams with high fidelity in Stackelberg games.

Given the described research gaps, the contributions of this paper are threefold:
\begin{enumerate}
    \item We introduce the novel $w$-PPSM which allows for the sharing of differentially-private data streams in Stackelberg games with high fidelity. This mechanism uses an optimization-based approach to recover the fidelity and feasibility of the obfuscated data w.r.t. the original Stackelberg game. This mechanism is developed for the target application of the coordination between heat and electricity markets and the exchange of hourly electricity loads over a 24-hour window.
    \item We show that the $w$-PPSM satisfies interesting theoretical properties. In particular, it achieves strong privacy goals while providing a bound on the error introduced on the obfuscated sensitive data.
    \item Through multiple numerical simulations, we show the efficiency and robustness of the $w$-PPSM under varying privacy parameters and operating conditions. The numerical results show that the $w$-PPSM can achieve up to two orders of magnitude cost reduction compared to a standard differentially-private mechanism.
\end{enumerate}

The remainder of this paper is organized as follows. Section \ref{sec:coordination} introduces the target application, Section \ref{sec:DP} summarizes the background on DP, Section \ref{sec:PPSM} defines the proposed $w$-PPSM, Section \ref{sec:numerical} presents the numerical results, and Section \ref{sec:conclusion} concludes this paper.

\begin{small}
\makenomenclature
\renewcommand\nomgroup[1]{%
  \item[\bfseries
  \ifstrequal{#1}{A}{A. Leader and follower's data ($D^{\text{L}}$, $D^{\text{F}}$)}{%
  \ifstrequal{#1}{B}{B. Leader and follower's variables}{}
  }
]}
\nomenclature[A]{$\text{L}^\text{H}_{lt}$}{Heat load $l$ at time $t$ (Wh)} \nomenclature[A]{$\text{L}^\text{E}_{lt}$}{Electricity load $l$ at time $t$ (Wh)} 
\nomenclature[A]{$\text{H}^\text{min}_{jt}$}{Minimum heat output of supplier $j$ at time $t$ (Wh)}
\nomenclature[A]{$\text{H}^\text{max}_{jt}$}{Maximum heat output of supplier $j$ at time $t$ (Wh)}
\nomenclature[A]{$\text{E}^\text{min}_{jt}$}{Minimum electricity output of supplier $j$ at time $t$ (Wh)}
\nomenclature[A]{$\text{E}^\text{max}_{jt}$}{Maximum electricity output of supplier $j$ at time $t$ (Wh)}
\nomenclature[A]{$\text{F}^\text{max}_j$ }{Maximum fuel consumption of CHP $j$ (Wh)} 
\nomenclature[A]{$\rho^{\text{E}}_j$}{Electricity efficiency ratio of CHP $j$ (-)} 
\nomenclature[A]{$\rho^{\text{H}}_j$}{Heat efficiency ratio of CHP $j$ (-)}
\nomenclature[A]{$\text{R}_j$}{Minimum power-to-heat ratio of CHP $j$ (-)} 
\nomenclature[A]{$\text{COP}_j$}{Coefficient of performance of HP $j$ (-)}
\nomenclature[A]{$\text{C}^\text{H}_{jt}$}{Variable heat cost of supplier $j$ at time $t$ (\euro/Wh)}
\nomenclature[A]{$\text{TC}^\text{min}_{zz't}$}{Minimum transmission capacity from zone $z$ to $z'$ at time $t$ (Wh)}
\nomenclature[A]{$\text{TC}^\text{max}_{zz't}$}{Maximum transmission capacity from zone $z$ to $z'$ at time $t$ (Wh)}
\nomenclature[A]{$\text{C}^{\text{E}}_{jt}$}{Variable electricity cost of supplier $j$ at time $t$ (\euro/Wh)}
\nomenclature[B]{$\bm{h_{jt}}$}{Electricity production of supplier $j$ at time $t$ (Wh)}
\nomenclature[B]{$\bm{f_{zz't}}$}{Electricity flow from zone $z$ to $z'$ at time $t$ (Wh)}
\nomenclature[B]{$\bm{e_{jt}}$}{Electricity production of supplier $j$ at time $t$ (Wh)}
\nomenclature[B]{$\bm{\lambda^\textbf{E}_{zt}}$}{Electricity market price in zone $z$ at time $t$ (Wh)}
\nomenclature[B]{$\bm{e^\textbf{min}_{jt}}$}{Minimum electricity output of CHP or HP $j$ at time $t$ (Wh)}
\nomenclature[B]{$\bm{e^\textbf{max}_{jt}}$}{Maximum electricity output of CHP or HP $j$ at time $t$ (Wh)}
\printnomenclature
\end{small}

\section{Heat and Electricity Market Coordination} \label{sec:coordination}


\subsection{Interactions between Heat and Electricity Sectors}

In Nordic countries, heat and electricity systems are operated by sequential and independent competitive markets. The day-ahead heat market is traditionally cleared \textit{before} the day-ahead electricity market. In each day-ahead energy market, suppliers place price-quantity bids for each hour of the following day that are dispatched based on a merit-order and least-cost principle. The sequential\footnote{CHPs and HPs must place their bids in the heat market before the electricity market. And once the heat market has been cleared, they place their bids in the electricity market.} participation of combined heat and power plants (CHPs) and heat pumps (HPs) in both heat and electricity markets creates implicit interactions between the systems.

Firstly, the physical characteristics of CHPs and HPs induce a strong linkage between heat and electricity production. As a result, in the current day-ahead electricity market, the minimum and maximum electricity outputs of CHPs and HPs are defined by their day-ahead heat dispatch. This heat-driven approach limits the operational flexibility of these units in the electricity market, which may limit the penetration of renewable energy sources and increase electricity prices.

Additionally, the production costs of CHPs and HPs are intrinsically linked to their heat and electricity outputs. Indeed, the heat production cost $\Gamma^{\text{H}}_{j}$ of HPs represents the cost of purchasing electricity in the day-ahead market. Similarly, the heat production cost of CHPs represents their total production cost minus revenues from electricity sales. However, the current market framework does not account for the impact of the heat production of CHPs and HPs on the electricity market prices, which in turn, impact the production costs in the heat market and may result in an inefficient dispatch.

\subsection{Electricity-Aware Heat Market Framework}

This paper provides an extension of the EAHM developed in \cite{mitridati2020heat}. This market framework aims at improving the coordination between heat and electricity sectors by better accounting for the interactions between them, while maintaining the sequential order of their decisions. This coordination framework is a classic Stackelberg game, in which the decisions of the first player (leader) impact the decisions of the second player (follower), which, in turn turn, impact the objective of the leader. As illustrated in the upper-part of Fig. \ref{fig:5}, in the EAHM, the heat market operator (leader) tries to minimize heat production costs while anticipating the impact of the heat dispatch of CHPs and HPs on the electricity market outcomes, specifically on electricity prices, which in turn impact heat production costs. This EAHM can be modelled as a bilevel optimization problem, in which the the upper-level problem, representing the heat market clearing, is constrained by the lower-level problem, representing the electricity market clearing for a given value of the heat market outcomes (namely the minimum and maximum electricity outputs of CHPs and HPs). Hence, the lower-level problem $\mathcal{P}^\text{F} \left( \bm{e^\textbf{min}_{jt}},\bm{e^\textbf{max}_{jt}} , D^\text{F}\right)$, is formulated as:

\begin{small}
\begin{subequations}
	\begin{alignat}{2}
& \min_{\underset{\bm{f_{zz't}}}{\bm{e_{jt},}}}&& \sum_{t \in \mathcal{T}} \sum_{j \in \mathcal{J}^{\text{E}}} \text{C}_{jt}^{\text{E}} \bm{e_{jt}} \label{Eq. (F-0)} \\
&  \text{s.t.} && \sum_{l \in \mathcal{L}_z^{\text{E}}} L^\text{E}_{lt} = \sum_{j \in \mathcal{J}_z^{\text{E}}}  \bm{e_{jt}}  + \sum_{z' \in \mathcal{Z}^\text{E}}  \bm{f_{zz't}} : \bm{\lambda_{zt}^{\textbf{E}}} , \ \forall z \in \mathcal{Z}^\text{E} , t \in \mathcal{T}  \label{Eq. (F-1)}  \\
&  \quad && \text{\small{TC}}_{zz't}^\text{min} \leq  \bm{f_{zz't}} = -\bm{f_{z'zt}} \leq \text{\small{TC}}_{zz't}^\text{max} , \ \forall z , z' \in \mathcal{Z}^\text{E} , t \in \mathcal{T}  \label{Eq. (F-2)}  \\
& \quad && \text{E}^\text{min}_{jt} \leq  \bm{e_{jt}} \leq  \text{E}^\text{max}_{jt} , \ \forall j \in \mathcal{J}^{\text{E} \setminus \{\text{CHP} \cup \text{HP} \} },t \in \mathcal{T}   \label{Eq. (F-4a)-(F-4b)} \\
& \quad && \bm{e^\textbf{min}_{jt}} \leq  \bm{e_{jt}} \leq  \bm{e^\textbf{max}_{jt}} \  , \ \forall j \in \mathcal{J}^{ \text{CHP} \cup \text{HP}},t \in \mathcal{T} \label{Eq. (F-5a)-(F-5b)}
\end{alignat}
\end{subequations}
\end{small}

\noindent where \eqref{Eq. (F-0)} represents the electricity production cost, \eqref{Eq. (F-1)} is the electricity balance equation in each market zone, \eqref{Eq. (F-4a)-(F-4b)}  and \eqref{Eq. (F-5a)-(F-5b)} represent the electricity production (or consumption) bounds of electricity-only producers, as well as CHPs and HPs, respectively. Note that the bounds in \eqref{Eq. (F-5a)-(F-5b)} are decisions variables of the upper-level problem, and treated as input in the lower-level problem.

Additionally, the upper-level problem $\mathcal{P}^\text{L} \left(D^\text{L},D^\text{F}\right)$ is formulated as:

\begin{small}
\begin{subequations}
\begin{alignat}{2}
& \min_{\underset{\bm{e^\textbf{max}_{jt}} , \bm{e_{jt}} , \bm{\lambda_{zt}^\textbf{E}}}{\bm{h_{jt}}, \bm{e^\textbf{min}_{jt}},}  } && \sum_{ z \in \mathcal{Z}^\text{E}, t \in \mathcal{T}} \big[ \sum_{j \in \mathcal{J}_z^\text{H}} \text{C}_{jt}^{\text{H}} \bm{h_{jt}} - \sum_{j \in \mathcal{J}_z^\text{CHP}} (\bm{\lambda^\textbf{E}_{zt}} - \text{C}_{jt}^{\text{E}})\bm{e_{jt}}  \nonumber \\
& \quad &&  + \sum_{j \in \mathcal{J}_z^\text{HP}}  \dfrac{\bm{\lambda^\textbf{E}_{zt}}}{\text{COP}_j} \bm{h_{jt}} \big] \label{Eq. (L-0)} \\
& \text{s.t.} \ &&  \sum_{l \in \mathcal{L}_z^\text{H}} L_{lt} = \sum_{j \in \mathcal{J}_z^{\text{H}}}  \bm{h_{jt}}  , \ \forall z \in \mathcal{Z}^\text{H} ,  t \in \mathcal{T} \label{Eq. (L-1)} \\
& \quad && \text{H}^\text{min}_{jt} \leq  \bm{h_{jt}} \leq  \text{H}^\text{max}_{jt} \  , \ \forall j \in \mathcal{J}^{\text{H}},t \in \mathcal{T}  \label{Eq. (L-4a)-(L-4b)} \\
& \quad && \bm{e^\textbf{min}_{jt}} =  \bm{e^\textbf{max}_{jt}} =  - \dfrac{\bm{h_{jt}}}{\text{COP}_j} \  , \ \forall j \in \mathcal{J}^{\text{HP}} , t \in \mathcal{T} \label{Eq. (L-9a)} \\
& \quad && \bm{e^\textbf{min}_{jt}} = \dfrac{\bm{h_{jt}}}{\text{R}_j} \  , \ \forall j \in \mathcal{J}^{\text{CHP}} , t \in \mathcal{T} \label{Eq. (L-9b)} \\
& \quad && \bm{e^\textbf{max}_{jt}} = \dfrac{\text{F}^\text{max}_j - \rho^\text{H}_j \bm{h_{jt}}}{\rho^{\text{E}}_j } \  , \ \forall j \in \mathcal{J}^{\text{CHP}} , t \in \mathcal{T}  \label{Eq. (L-9c)} \\
& \quad && \{ \bm{e_{jt}} , \bm{\lambda_{zt}^\textbf{E}}  \} \in \text{sol. of } \mathcal{P}^\text{F} \left(\bm{e^\textbf{min}_{jt}} , \bm{e^\textbf{max}_{jt}} , D^\text{F}\right) \label{Eq. (L-10)},
\end{alignat}
\end{subequations}
\end{small}

\noindent where \eqref{Eq. (L-0)} represents the heat production cost as a function of electricity prices, \eqref{Eq. (L-1)} is the heat balance equation in each market zone\footnote{Each heat market zone represents a geographically isolated district heating network.}, \eqref{Eq. (L-4a)-(L-4b)} represents the heat production bounds for all heat suppliers, \eqref{Eq. (L-9a)}-\eqref{Eq. (L-9c)} define the minimum and maximum electricity production (or consumption) of CHPs and HPs, and \eqref{Eq. (L-10)} sets the electricity dispatch and prices as the optimal solutions of the lower-level problem. A detailed formulation of this bilevel optimization problem and its solution method is provided in \cite{mitridati2020heat}.

\section{Differential Privacy Framework} \label{sec:DP}

\subsection{Privacy Goals}

To achieve coordination, the leader problem takes the follower's data $D^\text{F}$ as input. This data includes price-quantity bids of suppliers and electricity loads of consumers for each hour of the following day. In line with recent regulatory changes, encouraging information exchange for sector coordination \cite{FERCreport}, the electricity suppliers' bids are considered as available information ($D^\text{F,a}$) shared with the leader. However, the loads of individual consumers is sensitive information ($D^\text{F,p}$) that needs to be protected to avoid the leakage of identifying or competitive information \cite{lisovich2008privacy}. Although the leader solely requires the aggregate electricity load in each market zone as input data, aggregation has been showed to be insufficient to protect individuals' data \cite{not_enough}. Therefore, in this target application, the framework of DP is applied to obfuscate the aggregate electricity loads in each market zone $z \in \mathcal{Z}$ over a 24-hour period, before sharing it with the leader.

For this purpose, the infinite sequence of hourly aggregate electricity loads is represented as so-called \textit{data stream} $L^{\text{E}}_{zt} = \sum_{j \in \mathcal{L}_z^{\text{E}}} L^{\text{E}}_{jt}$, with the tuples $(z,t)$ in the universe $\mathcal{U}=\mathcal{Z}x\mathcal{T}^{\infty}$, where $\mathcal{Z} =  \{1,...,Z\}$ is the set of users (market zones) and $\mathcal{T}^{\infty} = \{1,2,...\}$ is an unbounded set of time steps (hours)  \cite{OptStream}. In this data stream, an \textit{event} $L^{\text{E}}_t = \left[L^{\text{E}}_{1t},...,L^{\text{E}}_{Zt}\right]$ is defined as all the data points reported by the users that occurred at time $t$. The goal is therefore to apply DP to this data stream.

\subsection{Differential Privacy for Data Streams}

DP is a rigorous privacy notion which relies on the injection
of carefully calibrated noise to protect disclosures of the users' data, while allowing to extract information about the population \cite{dwork:13}. This framework enjoys several important properties, including \emph{composability} and \emph{immunity to post-processing}.
This paper adopts the \textit{$w$-privacy} framework \cite{W_levelP}, which extends the standard definition of DP to protect data streams within a time window of $w$ time steps. This framework operates on \textit{stream prefixes}, i.e., the sequence $L^{\text{E}}[t] = \left[L^{\text{E}}_1,...,L^{\text{E}}_t\right]$ of all events that occurred at or before time $t$. It relies on the notion of \textit{$w$-adjacency} \cite{OptStream} to capture the differential information to be protected, as defined below:
\begin{definition}
Two data streams prefixes $L^{\text{E}}[t] $ and $L'^{\text{E}}[t] $ are $w$-neighbors, denoted by $L^{\text{E}}[t]  \sim_w L'^{\text{E}}[t] $, if
\begin{enumerate}
    \item their elements (events) are pairwise neighbors, i.e. they differ at most by one element. This is formally defined for a given pair of events $L^{\text{E}}_i$ and $L'^{\text{E}}_i$, where $ i \in [t]$, as:  $\exists z$ s.t. $|L^{\text{E}}_{zi} - L'^{\text{E}}_{zi}| \leq \alpha $ and  $\forall z' \neq z$, $L^{\text{E}}_{z'i} = L'^{\text{E}}_{z'i}$ with $\alpha \in \mathbb{R}^+$ the indistinguishability parameter representing how much data variation has to be protected; and
    \item all the differing elements are within a time window of up to $w$ time steps. This is formally defined as: for any given $ i<j \in [t]$, if $L^{\text{E}}_i \neq L'^{\text{E}}_i$ and $L^{\text{E}}_j \neq L'^{\text{E}}_j$, then it holds that $j-i+1 \leq w$.
\end{enumerate}
\end{definition}

In the context of the target application, a mechanism is said to satisfy $w$-event $\epsilon$-differential privacy ($w$-privacy for short) if it satisfies the following definition: 
\begin{definition}
Let $\cM$ be a randomized algorithm that takes as input a stream prefix of 
arbitrary size and outputs an element from a set of possible output sequences $\cS$. Algorithm 
$\cM$ satisfies $w$-privacy if, for all $w$-neighboring stream prefixes $L^{\text{E}}[t] \sim_w L'^{\text{E}}[t]$, with $t \in \mathcal{T}^\infty$, and all sets $S \subseteq \cS$, it satisfies:

\begin{small}
\begin{equation}
  \frac{\mathbb{P}\left(\cM(L^{\text{E}}[t]) \in S\right)}{\mathbb{P}\left(\cM(L'^{\text{E}}[t]) \in S\right)} \leq exp(\epsilon),
\end{equation}
\end{small}

\noindent where 
$\epsilon \in \mathbb{R}^+$ is the \textit{privacy budget}.
\end{definition}

\subsection{Laplace Mechanism} \label{sec:laplace}

A commonly used method to achieve $w$-privacy for data streams is the so-called \textit{Laplace mechanism}. In the target application of this work, the privacy goal is to protect a data stream of aggregate loads within a 24-hour window. Therefore, we consider the Laplace mechanism $\mathcal{M}^{\text{Lap}}$ which takes as input a stream prefix $L^{\text{E}}[t]$ and outputs the sequence $\tilde{L}^{\text{E}}[t] = \left[\tilde{L}^{\text{E}}_1,...,\tilde{L}^{\text{E}}_t\right]$, such that $\tilde{L}^{\text{E}}_i = L^{\text{E}}_i + \xi_{i}$ where $\xi_{i} \in \mathbb{R}^Z$ is drawn from the i.i.d. Laplace distribution $Lap(\frac{w \alpha}{\epsilon})^Z$ for $i \in [t]$, with the time window parameter $w=24$. It is a well-known result that this Laplace mechanism achieves $w$-privacy with $w=24$ \cite{OptStream}.

The main limitation of this mechanism is that the original data is highly perturbed and the outcome of the algorithm is a data stream that, used as input to an optimization problem, may lead to severe fidelity and feasibility issues \cite{fioretto2021differential}. The $w$-PPSM introduced in this paper specifically aims at mitigating this issue.

\section{\emph{w}-Privacy-Preserving Stackelberg Mechanism} \label{sec:PPSM}

The PPSM developed in \cite{fioretto2021differential} allows the exchange of differentially private data of high fidelity between the agents of a Stackelberg game. This section describes an extension of the PPSM that achieves $w$-privacy for a data stream. Similarly to \cite{fioretto2021differential}, this paper assumes that the leader and the follower each have access to their own accurate prediction models ($\mathcal{M}^\text{L}$ and $\mathcal{M}^\text{F}$) that can privately forecast electricity market costs and prices. This assumption is realistic in energy systems, since prediction models are commonly used to efficiently bid in the markets.

\subsection{Steps}

\begin{figure}[!ht]
\vspace{-5pt}
\centering
\includegraphics[width=.99 \linewidth]{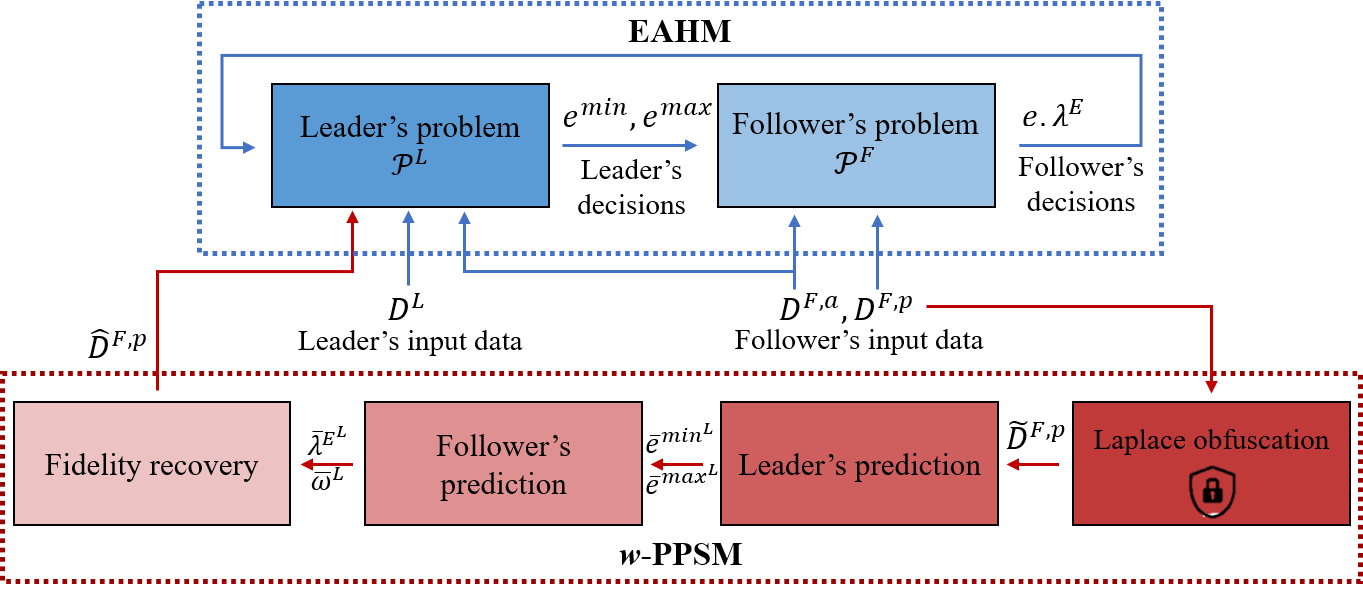}
\caption{EAHM and $w$-PPSM flowchart}
\label{fig:5}
\vspace{-5pt}
\end{figure}

The proposed $w$-PPSM ($\mathcal{M}^{\text{PPSM}}$) is performed each day, before the heat and electricity markets are cleared, to protect the sensitive data of the follower $D^{\text{F,p}}$ for each hour of the following day. The outcome of this mechanism is the privacy-preserving data $\hat{D}^{\text{F,p}}$ to be shared with the leader. The steps of this mechanism are schematically represented in Figure \ref{fig:5} and summarized below.

\subsubsection{\textbf{Laplace-obfuscation}} Firstly, the follower obfuscates the sensitive data ${D}^\text{F,p}$ according to the $w$-private Laplace mechanism $\mathcal{M}^{\text{Lap}}$ described in Section \ref{sec:laplace}, before releasing it to the leader.

\subsubsection{\textbf{Leader's prediction}} Using publicly available data and the Laplace-obfuscated data $\tilde{D}^\text{F,p}$ obtained in step (1), the leader estimates the values of the minimum and maximum electricity outputs of CHPs and HPs ($\bar{e}_{jt}^{\text{min}^{\text{L}}}$ and $\bar{e}_{jt}^{\text{max}^{\text{L}}}$) for the following day. To do so, it uses its prediction model $\mathcal{M}^{\text{L}}$ to predict the electricity prices $\bar{\lambda}^{\text{E}^{\text{L}}}_{zt}$. The leader then solves a decoupled heat market $\bar{\mathcal{P}}(D^{\text{L}},\bar{\lambda}^{\text{E}^{\text{L}}}_{zt})$, in which the follower's variables $\bm{\lambda^\text{E}_{zt}}$ in the objective function \eqref{Eq. (L-0)} are replaced by the predicted values $\bar{\lambda}^{\text{E}^{\text{L}}}_{zt}$, and the lower-level problem \eqref{Eq. (L-10)} is replaced by constraint \eqref{Eq. (F-5a)-(F-5b)} with $\bm{e}_{jt}$ a free variable. Note that this optimization problem does not take the follower's data $D^{\text{F}}$ as input. The solutions of this optimization problem ($\bar{e}_{jt}^{\text{max}^{\text{L}}}$ and $\bar{e}_{jt}^{\text{min}^{\text{L}}}$) are shared with the follower.

\subsubsection{\textbf{Follower's prediction}} With publicly available information, the Laplace-obfuscated data obtained in step (1), and the predicted values obtained in step (2), the follower predicts the electricity market costs $\bar{\omega}^{\text{F}}$ and prices $\bar{\lambda}^{\text{E}^{\text{F}}}_{zt}$ using its prediction model $\mathcal{M}^{\text{F}}$.

\subsubsection{\textbf{Fidelity recovery}} Given its own available data, the obfuscated data obtained in step (1) and the predicted values computed in steps (2) and (3), the follower derives the new privacy-preserving data $\bm{\hat{D}^{\text{F,p}}}$. To do so, it uses an optimization-based approach to optimally redistribute the noise on the sensitive data introduced in step (1) while recovering feasibility and fidelity w.r.t to the solutions of the original Stackelberg game. This bilevel optimization problem, inspired by \cite{fioretto2021differential}, is formulated as:

\begin{small}
\begin{subequations}
\begin{alignat}{2}
& \min_{\bm{\hat{D}^\textbf{F,p}}, \bm{\hat{\lambda}^{\textbf{E}}_{zt}}, \bm{\hat{\omega}^{\textbf{F}}}} &&
  \| \bm{\hat{D}^\textbf{F,p}} - \tilde{D}^\text{F,p}\|^2_2  \label{Eq. (PPSM-0)}\\
& \textbf{s.t.} && 
  |  \bm{\hat{\omega}^{\textbf{F}}} - \bar{\omega}^{F} | \leq \eta_p \label{Eq. (PPSM-1)} \\
& \quad && |  \bm{\hat{\lambda}^{\textbf{E}}_{zt}} - \bar{\lambda}^{\text{E}^{\text{F}}}_{zt} | \leq \eta_d , \forall z \in \mathcal{Z}, t \in \mathcal{T} \label{Eq. (PPSM-2)} \\
& \quad && \bm{\hat{\lambda}^{\textbf{E}}_{zt}} = \text{sol. of }
\mathcal{P}^\text{F}\left(\bar{e}_{jt}^{\text{min}^\text{L}},\bar{e}_{jt}^{\text{max}^\text{L}}, D^\text{F,a}, \bm{\hat{D}^\textbf{F,p}} \right), \label{Eq. (PPSM-3)}
\end{alignat}
\end{subequations}
\end{small}

\noindent where the objective \eqref{Eq. (PPSM-0)} is to find a vector of privacy-preserving data $\bm{\hat{D}^\textbf{F,p}}$ that minimizes the distance w.r.t. the Laplace-obfuscated data $\tilde{D}^\text{F,p}$, subject to fidelity constraints w.r.t. the predicted objective value $\bar{\omega}^{\text{F}}$ \eqref{Eq. (PPSM-1)} and electricity prices $\bar{\lambda}^{\text{E}^\text{F}}_{zt}$ \eqref{Eq. (PPSM-1)}, and feasibility constraints w.r.t. the follower's problem $\mathcal{P}^\text{F}\left(\bar{e}_{jt}^{\text{min}^\text{L}},\bar{e}_{jt}^{\text{max}^\text{L}}, D^\text{F,a}, \bm{\hat{D}^\textbf{F,p}} \right)$ in \eqref{Eq. (PPSM-3)}.
$\eta_p$ and $\eta_d$ are parameters specifying the desired
fidelity levels. Note that since the dual variables of the follower directly impact the leader's problem, \eqref{Eq. (PPSM-2)} indirectly enforces fidelity w.r.t. the leader's objective value. Furthermore, the follower's objective function $\hat{\omega}^{\text{F}}$ and dual variables $\hat{\lambda}^{\text{E}}_{zt}$ are defined as the solutions to the lower-level problem \eqref{Eq. (PPSM-3)}. The solutions to this optimization problem $\hat{D}^\text{F,p}$ are shared with the leader.

After the $w$-PPSM has been performed, the leader uses the privacy-preserving data $\hat{D}^{\text{F,p}}$ as input to solve its bilevel optimization problem $\mathcal{P}^\text{L}(D^{\text{L}},D^{\text{F,a}},\hat{D}^{\text{F,p}})$ described by \eqref{Eq. (L-0)}-\eqref{Eq. (L-10)}.

\subsection{Theoretical Properties}

A direct extension of \cite{fioretto2021differential} ensures that the proposed $w$-PPSM satisfies important theoretical properties, among which, the most important are:
\begin{enumerate}
    \item \textit{Privacy}: For given positive real values of the parameters $\alpha$, $\epsilon$, $\eta_p$ and $\eta_d$, the proposed $w$-PPSM mechanism satisfies $w$-privacy.
    \item \textit{Error on sensitive data}: After the fidelity-recovery phase, the expected error induced by the $w$-PPSM on the original sensitive data is bounded by the inequality: $\mathbb{E}[ \| \hat{D}^{\text{F,p}} - D^{\text{F,p}} \| ] \leq 4 (w\alpha)^2$.
\end{enumerate}
The first property can intuitively be justified by the immunity to post-processing of the Laplace mechanism in step (1), and the fact that all subsequent steps (2)-(4) do not access the original sensitive data. The second property is derived using triangular inequalities.

\section{Numerical Results} \label{sec:numerical}

This numerical analysis evaluates the performance of the $w$-PPSM in comparison to the Laplace mechanism.

\subsection{Case Study Setup}

The case study considered is a simplified version of the one used in \cite{mitridati2019bid}, which represents a modified version of the IEEE 24-bus system coupled with two 3-node district heating networks, in which network constraints are neglected. The overall system consists of four CHPs, two HPs, four heat-only generators, two heat storage units, twelve synchronous electricity generators, and six wind farms. Heat and electricity system parameters, as well as time series of heat and electricity loads and wind power generation for a given day
are derived from \cite{ordoudis2016updated,mitridati2019bid,energinet} and available in the online appendix \cite{online_appendix_1}.

For this case study, the privacy budget $\epsilon$ is fixed to $1$, and the fidelity parameters $\eta_p$ and $\eta_d$ are fixed to $0.1\%$ of the follower's objective and $10.0\%$ of the electricity prices, respectively. All the values displayed are average results over several instances.

\subsection{Results}

Table \ref{tab:1} reports the error on the original sensitive data, and the leader and follower's \textit{costs of privacy}, defined as the relative errors on the objective values of the leader and the follower, achieved by the Laplace mechanism and the $w$-PPSM for different values of the indistinguishability parameter $\alpha$\footnote{The chosen values of $\alpha$ guarantee a low privacy risk since the aggregate electricity demand ranges between $644.47$MWh and $2498.54$MWh.}, which represents how much variation of load is protected. As expected, since the parameter $\alpha$ determines the level of noise added to the original data, the errors on the sensitive data and the leader's cost of privacy induced by the Laplace mechanism drastically increase as $\alpha$ grows. On the contrary, the $w$-PPSM shows substantially better performances, and these errors remain stable with the increase of the parameter $\alpha$. For larger values of $\alpha$  ($\geq 50$), the $w$-PPSM achieves up to one order of magnitude reduction in the error on the sensitive data, and two orders of magnitude reduction in the leader's cost of privacy.

We also observe that the follower's cost of privacy, for both mechanisms, slightly decreases with higher values of $\alpha$. Intuitively, this can be explained by the interactions between the leader and the follower in the Stackelberg game. As the noise added to the electricity demand increases, the leader is less capable of anticipating the reaction of the follower, and of optimizing its own objective at the expense of the follower. Similar observations have been made related to the impact of DP on truthfulness in mechanism design \cite{mcsherry:07}. Furthermore, the $w$-PPSM consistently achieves better performances compared to the Laplace mechanism, and up to two orders of magnitude reduction in the follower's cost of privacy.

\begin{table}[!ht]
\vspace{-5pt}
\centering
\caption{Errors on the electricity demand vector ($\Delta_{D^{\text{F,p}}}$), objective values of the leader ($\Delta_{\omega^\text{L}}$) and the follower ($\Delta_{\omega^\text{F}}$) for varying indistinguishability parameters $\alpha$, averaged over $100$ instances.} \label{tab:1}
\resizebox{0.9\linewidth}{!}
{
\begin{tabular}{ll @{\hspace{5pt}}|@{\hspace{5pt}} rrr}
$\bm{\cM}$ & $\bm{\alpha}$ & $\bm{\Delta_{D^{\textbf{F,p}}}}$ \textbf{(L1)} & $\bm{\Delta_{\omega^\textbf{L}} (\%)}$ &   $\bm{\Delta_{\omega^\textbf{F}} (\%)}$ \\
\textbf{Laplace }      & 10.0    & 6139.88 & 0.764773 & 8.751309  \\
       & 50.0    & 34131.08 & 47.556005 & 6.352331 \\
       & 100.0  & 39131.19 & 58.455686 &   5.430761 \\
\hline
\textbf{PPSM}      & 10.0   &  3723.66 & 0.842956 &  1.067518  \\
       & 50.0   & 3843.56 & 0.606088 & 0.483239 \\
       & 100.0   & 3296.58  & 0.302367 &   0.058785 \\
\end{tabular}
}
\vspace{-5pt}
\end{table}

\begin{figure}[!ht]
\vspace{-5pt}
\centering
\includegraphics[width=1.0 \linewidth]{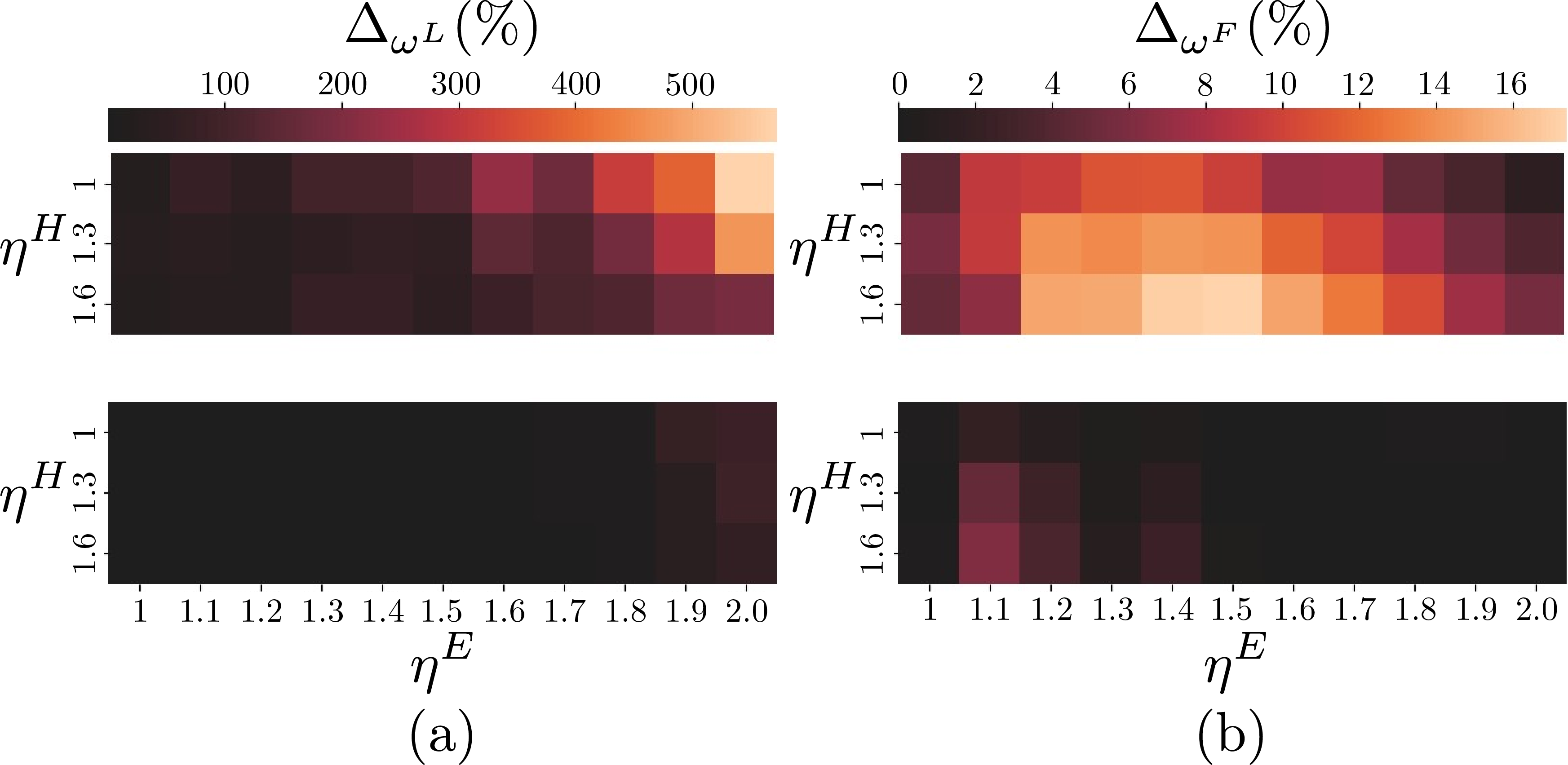}
\caption{(a) Leader's cost of privacy ($\Delta_{\omega^\text{L}} in \%$) at varying  heat ($\eta^\text{H}$) and electricity ($\eta^\text{E}$) stress levels, obtained via the Laplace mechanism (top) and $w$-PPSM (bottom). (a) Follower's cost of privacy ($\Delta_{\omega^\text{F}} in \%$) at varying  heat ($\eta^\text{H}$) and electricity ($\eta^\text{E}$) stress levels, obtained via the Laplace mechanism (top) and $w$-PPSM (bottom) for $\alpha\!=100.0$(MWh), averaged over 20 instances.}
\label{fig:1}
\vspace{-5pt}
\end{figure}

Figure \ref{fig:1} presents heat maps of the leader and follower's costs of privacy under varying operating conditions in both heat and electricity systems. These operating conditions in the heat (electricity) system are represented by the varying stress factors $\eta^\text{H}$ ($\eta^\text{E}$) representing the multiplying factors applied to the heat (electricity) loads of the reference day previously considered. In this analysis, the heat load is uniformly increased by $30\%$ to $60\%$, and the electricity load by $10\%$ to $100\%$.

Overall, this stress analysis underlines once more the robustness of the $w$-PPSM under various operating conditions. Indeed, the $w$-PPSM succeeds in keeping the leader and follower's cost of privacy very low compared to the Laplace mechanism, for all the stress factor levels. Under certain operating conditions, the $w$-PPSM achieves up to two orders of magnitude reduction in the leader and follower's costs of privacy. 

We also notice that the highest costs of privacy for each mechanism are achieved under different combinations of stress factors.
The Laplace mechanism performs especially poorly for the leader's cost of privacy for high values of the electricity stress factor. Intuitively, this can be explained by the fact that, for higher electricity loads, the volatility of the electricity prices is increased, which in turn, impacts the merit order in the heat market and leads to a sub-optimal dispatch. However, this error is somehow reduced for corresponding higher values of the heat stress factor. Indeed, with higher heat loads, the relative share of HPs and CHPs in the heat dispatch, and therefore their impact on the leader's objective value, decreases.
Furthermore, the Laplace mechanism achieves the highest follower's cost of privacy for the highest heat stress factor. Intuitively, this can be explained by the fact that, with higher heat loads, the heat dispatch of HPs and CHPs increases, which reduces their operational flexibility in the electricity market. These tightened interactions between heat and electricity markets result in higher errors on the electricity costs. 
This analysis identifies the system's operating conditions that are the most vulnerable to perturbations and the ones resulting in a negligible cost of privacy when applying DP. This information can be leveraged to reduce the privacy budget \cite{OptStream}.

\section{Conclusion} \label{sec:conclusion}

This paper introduces the $w$-PPSM which generates differentially-private data streams with high fidelity that can be used as input to the EAHM to coordinate the operation of heat and electricity systems.
The $w$-PPSM was shown to enjoy strong theoretical properties. Furthermore, the numerical results show that the $w$-PPSM achieves up to two orders of magnitude reduction in the costs of privacy in both heat and electricity systems compared to the traditional Laplace mechanism.

Future work will aim at developing theoretical bounds on the costs of privacy, and accounting for potential correlations between the users' data streams. Furthermore, focus will be placed on reducing the costs of privacy. Advanced obfuscation methods can be used to reduce the initial noise added to the data. And, the sparse vector technique can be adapted to privately identify the operating conditions resulting in negligible costs of privacy, and adapt the noise added under these conditions to reduce the privacy budget \cite{OptStream}.


\bibliographystyle{IEEEtran}
\bibliography{bibliography}

\begin{thebibliography}{10}
\providecommand{\url}[1]{#1}
\csname url@samestyle\endcsname
\providecommand{\newblock}{\relax}
\providecommand{\bibinfo}[2]{#2}
\providecommand{\BIBentrySTDinterwordspacing}{\spaceskip=0pt\relax}
\providecommand{\BIBentryALTinterwordstretchfactor}{4}
\providecommand{\BIBentryALTinterwordspacing}{\spaceskip=\fontdimen2\font plus
\BIBentryALTinterwordstretchfactor\fontdimen3\font minus
  \fontdimen4\font\relax}
\providecommand{\BIBforeignlanguage}[2]{{%
\expandafter\ifx\csname l@#1\endcsname\relax
\typeout{** WARNING: IEEEtran.bst: No hyphenation pattern has been}%
\typeout{** loaded for the language `#1'. Using the pattern for}%
\typeout{** the default language instead.}%
\else
\language=\csname l@#1\endcsname
\fi
#2}}
\providecommand{\BIBdecl}{\relax}
\BIBdecl

\bibitem{heatroadmap}
\BIBentryALTinterwordspacing
{Heat Road Map {E}urope}, ``A low-carbon heating and cooling strategy for
  {E}urope,'' 2018. [Online]. Available: \url{http://www.heatroadmap.eu/}
\BIBentrySTDinterwordspacing

\bibitem{pinson2017towards}
P.~Pinson, L.~Mitridati, C.~Ordoudis, and J.~Ostergaard, ``Towards fully
  renewable energy systems: Experience and trends in denmark,'' \emph{{CSEE} J.
  Power Energy Syst.}, vol.~3, no.~1, pp. 26--35, 2017.

\bibitem{mitridati2020heat}
L.~Mitridati, J.~Kazempour, and P.~Pinson, ``Heat and electricity market
  coordination: A scalable complementarity approach,'' \emph{Eur. J. Oper. Res.
  (EJOR)}, vol. 283, no.~3, pp. 1107--1123, 2020.

\bibitem{mitridati2019bid}
\BIBentryALTinterwordspacing
L.~Mitridati, P.~Van~Hentenryck \emph{et~al.} (2019) A bid-validity mechanism
  for sequential heat and electricity market clearing. [Online]. Available:
  \url{arXiv preprint arXiv:1910.08617}
\BIBentrySTDinterwordspacing

\bibitem{byeon2019unit}
G.~Byeon and P.~Van~Hentenryck, ``Unit commitment with gas network awareness,''
  \emph{{IEEE} Trans. Power Syst.}, vol.~35, no.~2, pp. 1327--1339, 2020.

\bibitem{FERCreport}
\BIBentryALTinterwordspacing
{{FERC} and {NERC}}, ``Staff report on outages and curtailments during the
  southwest cold weather event of february 1-5, 2011: Causes and
  recommendations,'' Aug. 2011. [Online]. Available:
  \url{https://www.nerc.com/pa/rrm/ea/Pages/September-2011-Southwest-Blackout-Event.aspx}
\BIBentrySTDinterwordspacing

\bibitem{lisovich2008privacy}
M.~Lisovich and S.~Wicker, ``Privacy concerns in upcoming residential and
  commercial demand-response systems,'' \emph{{IEEE} Proc. Power Syst.},
  vol.~1, no.~1, pp. 1--10, 2008.

\bibitem{maharjan2013dependable}
S.~Maharjan, Q.~Zhu, Y.~Zhang, S.~Gjessing, and T.~Basar, ``Dependable demand
  response management in the smart grid: A stackelberg game approach,''
  \emph{IEEE Trans. Smart Grid}, vol.~4, no.~1, pp. 120--132, 2013.

\bibitem{dwork:13}
C.~Dwork and A.~Roth, ``The algorithmic foundations of differential privacy,''
  \emph{Theor. Comput. Sci.}, vol.~9, no. 3-4, pp. 211--407, 2013.

\bibitem{W_levelP}
G.~Kellaris, S.~Papadopoulos, X.~Xiao, and D.~Papadias, ``Differentially
  private event sequences over infinite streams,'' \emph{Proc. {VLDB}
  Endowment}, vol.~7, no.~12, pp. 1155--1166, 2014.

\bibitem{OptStream}
F.~Fioretto and P.~Van~Hentenryck, ``Optstream: Releasing time series
  privately,'' \emph{J. Artif. Intell. Res. ({JAIR})}, vol.~65, 2019.

\bibitem{mak2019privacy}
T.~W. Mak, F.~Fioretto, L.~Shi, and P.~Van~Hentenryck, ``Privacy-preserving
  power system obfuscation: A bilevel optimization approach,'' \emph{{IEEE}
  Trans. Power Syst.}, vol.~35, no.~2, pp. 1627--1637, 2019.

\bibitem{fioretto2021differential}
F.~Fioretto, L.~Mitridati, and P.~{Van Hentenryck}, ``Differential privacy for
  {S}tackelberg games,'' in \emph{Proc. 29th Int. Joint Conf. Artif. Intell.
  ({IJCAI}-2020)}, Jan 2021.

\bibitem{not_enough}
N.~Buescher, S.~Boukoros, S.~Bauregger, and S.~Katzenbeisser, ``Two is not
  enough: Privacy assessment of aggregation schemes in smart metering.''
  \emph{Proc. Priv. Enhancing Technol.}, vol. 2017, no.~4, pp. 198--214, 2017.

\bibitem{ordoudis2016updated}
\BIBentryALTinterwordspacing
C.~Ordoudis, P.~Pinson, J.~M. Morales, and M.~Zugno. (2016) An updated version
  of the {IEEE} {RTS} 24-bus system for electricity market and power system
  operation studies. [Online]. Available:
  \url{http://orbit.dtu.dk/files/120568114/An}
\BIBentrySTDinterwordspacing

\bibitem{energinet}
\BIBentryALTinterwordspacing
Energinet.dk. (2020) Danish system operator market data. [Online]. Available:
  \url{https://en.energinet.dk/}
\BIBentrySTDinterwordspacing

\bibitem{online_appendix_1}
\BIBentryALTinterwordspacing
L.~Mitridati, P.~Van~Hentenryck, and J.~Kazempour. (2021, Nov.) Supplementary
  material - case study 1. [Online]. Available:
  \url{https://doi.org/10.5281/zenodo.5717239}
\BIBentrySTDinterwordspacing

\bibitem{mcsherry:07}
F.~McSherry and K.~Talwar, ``Mechanism design via differential privacy,'' in
  \emph{48th Annu. {IEEE} Symp. Found. Comput. Sci. ({FOCS'07})}.\hskip 1em
  plus 0.5em minus 0.4em\relax IEEE, 2007, pp. 94--103.

\end{thebibliography}

\end{document}